\documentclass[aps,twocolumn,showpacs,groupedaddress,preprintnumbers]{revtex4}

\usepackage{epsf}
\usepackage{graphicx}
\usepackage{dcolumn}
\usepackage{bm}
\usepackage{ulem}

\begin{document}
\bibliographystyle{apsrev}


 \flushbottom
 \renewcommand{\textfraction}{0}
 \renewcommand{\topfraction}{1}
 \renewcommand{\bottomfraction}{1}

\newcommand{\beq}{\begin{equation}}
\newcommand{\dd}{\partial}
\newcommand{\eeq}{\end{equation}}
\newcommand{\bea}{\begin{eqnarray}}
\newcommand{\eea}{\end{eqnarray}}



\preprint{EPFL-ITP-LPPC-04-2s} 
\preprint{UCLA/04/TEP/13}

\title{Supersymmetric dark matter Q-balls and their interactions in matter 
}

\author{Alexander Kusenko$^{1,2}$, Lee C. Loveridge$^1$, 
and Mikhail Shaposhnikov$^3$
}
\affiliation{$^1$Department of Physics and Astronomy, UCLA, Los Angeles, CA
90095-1547 \\ $^2$RIKEN BNL Research Center, Brookhaven National
Laboratory, Upton, NY 11973 \\
$^3$Institute of Theoretical Physics, Swiss Federal Institute
of Technology (Lausanne), BSP, 
CH-1015 Lausanne, Switzerland
}

%


\begin{abstract}
\vspace{0.1cm}

Supersymmetric extensions of the Standard Model contain non-topological
solitons, Q-balls, which can be stable and can be a form of cosmological
dark matter.  Understanding the interaction of SUSY Q-balls with matter
fermions is important for both astrophysical limits and laboratory searches
for these dark matter candidates.  We show that a baryon scattering off a
baryonic SUSY Q-ball can convert into its antiparticle with a high
probability, while the baryon number of the Q-ball is increased by two
units.  For a SUSY Q-ball interacting with matter, this process dominates
over those previously discussed in the literature.

\end{abstract}

\pacs{ 11.30.Pb 12.60.Jv 95.35.+d 14.80.Ly   \hspace{1.0cm} } 

\maketitle

\renewcommand{\thefootnote}{\arabic{footnote}}
\setcounter{footnote}{0}

\section{Introduction}

Supersymmetry (SUSY) is widely regarded as a plausible candidate for physics
beyond the Standard Model.  In addition to its aesthetic appeal and
its ability to stabilize the hierarchy of scales, supersymmetry
provides dark matter candidates in the form of the lightest
supersymmetric particles and SUSY Q-balls.

Q-balls are nontopological solitons that carry some conserved global
quantum number \cite{q,nts_review}. They exist in SUSY extensions of the
standard model~\cite{ak_mssm} and are either stable or have lifetimes in
excess of the age of the universe in theories with gauge-mediated SUSY
breaking~\cite{dks}. Stable SUSY Q-balls have large baryon numbers; the
vacuum expectation value (VEV) of the scalar field inside these Q-balls
lies along some ``flat direction'' of the effective potential.

Cosmological evolution of SUSY scalar fields after inflation can give
rise to both ordinary matter, via Affleck-Dine
baryogenesis~\cite{ad,dk}, and to dark matter in the universe in the
form of SUSY Q-balls \cite{ks} or neutralinos from Q-ball
decays~\cite{enqvist_review}. Since both atomic matter and dark
matter arise from the same process, it may be possible to relate
their amounts in the
universe~\cite{ls,fh_1,fh_2,fy_d,Banerjee:2000mb}. In theories with
gravity mediated SUSY breaking, all Q-balls are short-lived, but
their production can change the standard way of computing neutralino
abundances~\cite{em,fh_1,fh_2,fy_d}.

If baryonic SUSY Q-balls make up the dark matter in the universe,
they may be detected \cite{kkst} in current or future experiments,
such as ANTARES, Baikal, IceCube, Super-Kamiokande, {\it etc}. 
A number of bounds on Q-balls already exist in the
literature~\cite{kkst,exp}.

Specific experimental signatures, which could be used to detect SUSY
Q-balls and to distinguish them from other heavy dark-matter
candidates, such as strangelets, nuclearites, and monopoles, depend
on the Q-ball interactions in matter.  These interactions also determine
the astrophysical bounds, which will be discussed elsewhere~\cite{future}.
Baryonic Q-balls in gauge-mediated
scenarios have mass per baryon below 1~GeV.  Hence, storing baryon
number in Q-balls is energetically favored over nucleons.  It is clear
that a nucleon can be absorbed by a Q-ball and the energy difference
can be released.  However, the rates of such interactions have not
been computed accurately.  

It was conjectured in Ref.~\cite{kkst} that an absorption of a
nucleon by a Q-ball occurred essentially in two stages.  Since the
color SU(3) symmetry inside a baryonic SUSY Q-ball is
broken~\cite{kst}, it was concluded that the first stage is a nucleon
disintegration into quarks, in which the energy released (in pions)
is of the order of the binding energy of quarks, that is $\sim
1$~GeV.  The next step was a decay of quarks into the scalar
condensate inside the Q-ball. This process is, of course, allowed
because the energy per baryon number in the condensate is small.  The
rate of this second process was estimated to be suppressed by a
factor $(E/\Lambda)^3$, where $ \Lambda \sim 1$~TeV is the scale of
SUSY breaking and $E\sim 1$ GeV is the typical quark energy. This
estimate was essential for the discussion of Q-balls interactions in
a neutron star~\cite{sw}. On the basis of this estimate it was
concluded in Ref.~\cite{sw} that a neutron star can survive for at
least 1 Gyr after capturing a dark-matter Q-ball. We will see that
the picture of interaction of baryons with Q-balls discussed above is
incorrect.

The goal of the present paper is to reanalyze the interaction of
Q-balls with ordinary matter. We find, in fact, that quarks, falling
on a Q-ball, are reflected as antiquarks with a probability on the
order of one, practically independent of the parameters of the
theory. In other words, Q-balls convert the matter into antimatter on
their surface (or antimatter into matter, if placed in an anti-matter
environment). Baryon number is conserved during this process: after
reflection of an antinucleon, the baryonic charge of the Q-ball
increases by 2 units.

\section{Interactions of quarks with SUSY Q-balls}
\label{formalism}
\subsection{SUSY Q-ball basics}

Let us consider some flat direction of a supersymmetric extension of
the standard model.  This flat direction can be parameterized by a
scalar field $\varphi$.  The VEVs of squarks and sleptons are
proportional to $\varphi$ along the flat direction.  In theories with
gauge mediation the effective potential along this flat direction
practically does not grow $V(\varphi) \propto \Lambda^4$, up to some
$\varphi_{\rm max}$ which depends on the flat direction \footnote{We
omit possible $\log$ factors, because they are not essential for the
discussion}.  There exists a classical spherically symmetric solution
of the field equations, a non-topological soliton, in which 
\beq
\varphi = \phi_0 f(r)e^{i \omega t},  
\eeq 
where $f(r) \simeq \sin(\omega r)/(\omega r)$ for $r \leq R=\pi/\omega$ and
zero for $r>R$. It carries the global charge $Q$.  If
$\phi_0<\varphi_{\rm max}$, different parameters depend on $Q$ as
follows: 

\beq 
\phi_0 \sim \Lambda Q^{1/4},~~M \sim \Lambda Q^{3/4},~~
\omega \sim \Lambda Q^{-1/4}~.  
\label{fdprop}
\eeq

\subsection{Structure of the quark mass term inside Q-ball}

For us the relevant parts of the MSSM Lagrangian are those which describes 
interactions of quarks $\psi$ with squarks $\phi$ and gluinos $\lambda$:
\begin{equation}
{\cal L} =  -g {\sqrt{2}} T^{a}_{ij}(\lambda^{a} \sigma^2 \psi_j
\phi^{*}_i) + C.C.+... 
\label{Lmixingterm}
\end{equation}
and also the Majorana mass terms for gluinos: 
\beq
{\cal L_M} = M\lambda_a\lambda_a~.
\eeq
In theories with gauge mediation the Majorana gluino mass stays
constant along the flat directions of the effective potential
\cite{Giudice:1998bp}.

Inside a Q-ball the squarks have a non-zero expectation value $
\langle \phi \rangle = \varphi $, and, therefore, the quarks acquire a
non-zero mass through their mixing~(\ref{Lmixingterm}) with gluinos. 
In general, the mass matrix has many terms of the form 
\begin{equation}
M^{a}_{i}\lambda^{a} \sigma^2 \psi_i,
\end{equation}
where
\begin{equation}
M^{a}_{i}=-g {\sqrt{2}} T^{a}_{ji}\varphi^{*}_j~.
\end{equation}

To describe the interactions of quarks with the Q-ball, it is
sufficient to consider a simplified mixing matrix 
\begin{equation}
\left( \begin{array} {ccc}  0 & m & \varphi_L \\ m & 0 & \varphi_R \\
 		\varphi_L & \varphi_R & M  \end{array} \right)~,
\end{equation}
where $\varphi_{L,R}$ represent the mixing terms for the left and right
handed components of the quark field, and are proportional to the left and
right handed squark expectation values.  The factor of $- \sqrt{2} g$ has
also been absorbed into the definition of these parameters.  Since $\omega
\ll \phi_0$, we neglect the time dependence of the Q-ball solution for now.
It will be addressed below.  As a further simplification, one can neglect
the quark Dirac mass $m$, since along the MSSM flat directions the
Higgs field is equal to zero \cite{Gherghetta:1995}.  We
parameterize the left and right handed mixing terms as
\begin{equation}
\left( \begin{array} {c} \varphi_L \\ \varphi_R \end{array}\right) = 
\varphi\left( \begin{array} {c} \cos\alpha 
\\ \sin\alpha \end{array}\right)~.
\end{equation}
Then one linear combination of left and right handed quark fields 
\begin{equation}
\left( \begin{array} {c} \chi_L \\ \chi_R \end{array}\right) = 
\chi_{\rm massless}\left( \begin{array} {c} \sin\alpha 
\\ -\cos\alpha \end{array}\right)
\label{massless}
\end{equation}
remains massless.  The other, massive component
\begin{equation}
\left( \begin{array} {c} \chi_L \\ \chi_R \end{array}\right)= 
\chi_{\rm mass}\left( \begin{array} {c} \cos\alpha 
\\ \sin\alpha \end{array}\right)
\label{massive}
\end{equation}
is mixed with the gluino according to a mass matrix of the form
\begin{equation}
\left( \begin{array} {cc}  0 & \varphi \\
 		\varphi & M  \end{array} \right)~,
\end{equation}
where $\varphi = \sqrt{{\varphi_L}^2+{\varphi_R}^2}$.  This mass
matrix can be written as 
\begin{equation}
\tilde{M} \left( \begin{array} {cc} 0 & \sin{2\beta} \\
\sin{2\beta} & 2 \cos{2 \beta} \end{array}
\right)~.
\end{equation}
It has the eigenvalues 

\begin{equation} 
M_+ = 2 \tilde{M} \cos^2{\beta}, \ {\rm and} \ \ 
- M_- = - 2 \tilde{M} \sin^2{\beta} \ 
\end{equation}
corresponding to eigenvectors

\beq
\left(
\begin{array} {c}
\sin{\beta} \\ \cos{\beta}
\end{array}
\right)
\ {\rm and} \ 
\left(
\begin{array} {c}
\cos{\beta} \\ -\sin{\beta}
\end{array}
\right).
\label{simpeigen}
\end{equation}
These eigenstates form Majorana particles with masses $M_+,~(-M_-)$.

\subsection{Quark scattering on Q-balls}
Let us consider the quark scattering on a Q-ball. 
The Q-ball size is in general much larger than the wavelength of the 
quarks, so it is sufficient to treat the problem as reflection from
a  flat boundary surface. To simplify the analysis further, we will
assume that the profile of the scalar field has a $\theta$-function
behavior, which does not lead to any loss of accuracy since the
energy of the incident quark is much smaller than the masses
$M_+,~M_-$.

Then this is a simple quantum-mechanical problem in which one has to
match the the wave functions of the incident quark and the reflected
quark/antiquark (and, potentially, the gluino, if the energy of the
incoming quark is large enough) with the wave-functions of the
Majorana fermions inside the Q-ball. 

Majorana fermions obey the differential equation
\begin{equation}
i \bar{\sigma} \cdot \partial \chi + i m \sigma^2 \chi^* = 0
\end{equation}
with the solutions
\begin{equation}
\begin{array} {cc}
\sqrt{\sigma \cdot p} 
\left( A e^{-i p \cdot x} + \sigma^2 A^* e^{i p \cdot x} \right) & \ {\rm
for}\ 
m>0~, \\
\sqrt{\sigma \cdot p}
\left( A e^{-i p \cdot x} - \sigma^2 A^* e^{i p \cdot x} \right) & \ {\rm for}
\ m<0~. 
\end{array}
\end{equation}

Five different wave functions are involved as  summarized in Table
II. The combined wave functions of  the incoming  quark, the
reflected quark, and the reflected gluino must match the  internal
wave functions of the internal mass eigenstates at the boundary. 

\begin{table}
\begin{tabular}{|c|c|c|c|} \hline
Particle & Description & Mass & Direction \\ \hline
$A$ & Incoming Quark & $0$ & $+$ \\ \hline
$B$ & Reflected Quark & $0$ & $-$ \\ \hline
$C$ & Reflected Gluino & $M$ & $-$ \\ \hline
$D$ & Transmitted State & $+M_+$ & $+$ \\ \hline
$F$ & Transmitted State & $-M_-$ & $+$  \\ \hline
\end{tabular}
\label{quarkstates}
\caption{Different states participating in the scattering problem.}
\end{table}

The general continuity equation is 
\begin{equation}
\left( \begin{array} {c} \chi_{A}+\chi_{B} \\ \chi_{C} \end{array} \right)
= \left( \begin{array} {c} \sin{\beta} \\ \cos{\beta} \end{array} \right)
\chi_{D} +\left( \begin{array} {c} \cos{\beta} \\ -\sin{\beta} \end{array}
\right) \chi_{F}~,
\end{equation} 
which simplifies to {\small
\begin{eqnarray}
\sin{\beta}\sqrt{E}(A+B) + \cos{\beta} \sqrt{E+M} C& = &\sqrt{E+M_+}
D~,
\nonumber 
\\
\sin{\beta}\sqrt{E}({\bf \sigma \cdot \hat{p}_A} A 
           + {\bf \sigma \cdot \hat{p}_B} B)
\\ 
+ \cos{\beta}\sqrt{E-M} {\bf \sigma \cdot \hat{p_C}} C &=& \sqrt{E-M_+} 
{\bf \sigma \cdot \hat{p}_D} D~,
\nonumber
 \\
\cos{\beta}\sqrt{E}(A+B) - \sin{\beta} \sqrt{E+M} C& = & - \sqrt{E-M_-}
	{\bf \sigma \cdot \hat{p}_F} F~,
\nonumber
\\ 
\cos{\beta}\sqrt{E}({\bf \sigma
	\cdot \hat{p}_A} A + {\bf \sigma \cdot \hat{p}_B} B) 
\nonumber
\\ 
-\sin{\beta}\sqrt{E-M} {\bf \sigma \cdot \hat{p_C}} C &=&
	-\sqrt{E+M_+} F~.
	\nonumber
\end{eqnarray}}

In general, the gluino mass and the masses inside the Q-ball are much
larger than the energy of the incoming quark.  In this case $B=\pm i
A$ so the quark will be reflected with unit probability as it is
unable to propagate in any of the other states.  (The sign is
determined by the sign of the gluino mass.)

More interestingly, the reflected quark has reversed frequencies
compared  with the incoming quark.  This is because in changing the
direction, the effect of the projector $\sqrt{\sigma \cdot p}$ is
reversed.   Thus, if the quark comes in with a positive  frequency it
is reflected with a negative frequency.  This in turn  implies that
it must either change chirality or particle anti-particle  identity.
(A right-handed particle is described by its left-handed 
anti-particle, so the frequency flip may represent a change of
handedness.)  Zhitnitsky found similar results in studying reflections
from baryonic color superconductors \cite{Zhitnitsky:2004da}.

The linear combination of left and right handed quarks which remains 
massless will propagate through the Q-ball unimpeded and experiences
no  frequency flip.  However, because it mixes left and right handed
states,  it too can result in changes in handedness and particle
anti-particle  identity.

The exact probability of reflection, chirality change, or particle
identity change depends on the chirality and the identity of the
incoming particle and the left right composition of the squark
condensate.  These are summarized in Table II.  However, in a large
collection of many left- and right-handed quarks impinging on the
Q-ball, on average, half will be reflected and half will be
transmitted.  Half will change their identity and half will retain
their identity.

\begin{table}

Incoming Left Handed 

{\begin{tabular} {|c|c|c|}
\hline & Quark & Anti-Quark \\
\hline Transmitted & $\sin^4{\alpha}$ & $\sin^2{\alpha}\cos^2{\alpha}$ \\
\hline Reflected &   $\cos^2{\alpha}\sin^2{\alpha}$ & $\cos^4{\alpha}$ \\
\hline
\end{tabular}} \\
\begin{tabular}{l}
Total Probability to Reflect: $\cos^2 \alpha=({\varphi_L}^2/
{\varphi^2})$ \\ 
Total Probability to Change Identity: $\cos^2 \alpha({\varphi_L}^2/ {\varphi^2})$
 \end{tabular} 
\vspace{3mm} \\
Incoming Right Handed
{\begin{tabular} {|c|c|c|}
\hline & Quark & Anti-Quark \\
\hline Transmitted & $\cos^4{\alpha}$ & $\sin^2{\alpha}\cos^2{\alpha}$ \\
\hline Reflected &   $\cos^2{\alpha}\sin^2{\alpha}$ & $\sin^4{\alpha}$ \\
\hline
\end{tabular}} \\
\begin{tabular}{l}
Total Probability to Reflect: $\sin^2 \alpha=({\varphi_R}^2/
{\varphi^2})$ \\ 
Total Probability to Change Identity:  $\sin^2 \alpha({\varphi_R}^2/ {\varphi^2})$ 
\end{tabular} 
\label{probabilities}
\caption{The probabilities of a quark reflection from Q-ball as either a
  quark or an antiquark.}
\end{table}

The above results hold as long as all of the masses $M,~M_+,~M_-$ are
larger than the  energy.  Oblique reflection changes the formulas so
that $B=i \sigma^3 {\bf \sigma \cdot \hat{p}_A} A$, but a unitary
reflection and a frequency flip are preserved.  Also if the mass of
one of the internal  eigenstates is only slightly larger than the
incoming energy, the phase of the reflected state $B$ relative to $A$
changes from $i$ to $-1$, but again unit reflection and frequency
flipping are preserved.

However, if the gluino mass is much larger than the mixing terms,
one  of the two internal masses may be smaller than the energy of the
incoming  quark.  In this case the quark can propagate inside the
Q-ball, and the  reflection and frequency flip probabilities are
suppressed.  The reflected state is then 
$$
B=-\frac {E+M_r-p}{E+M_r+p} A~.
$$
In the case that the gluino mass is very large, so that $M_r=\frac
{\varphi^2} {M_g}\ll E$ we find that the reflection  probability goes
like ${|\frac {\varphi^2} {M_g E}|}^2$.  Since less of the  original
state is split off by reflection, the probability of becoming an 
anti-particle is also suppressed.  Thus, in this case the processing
of quarks by the Q-ball will be suppressed. This case, however, is not
realistic as the typical value of the gluino mass is assumed to be in
the range of hundreds GeV whereas the value of the squark fields
inside the Q-ball is  much higher. 

\subsection{Time-Dependent Effects}

Let us now consider the time dependence of the condensate.  The
squark condensate in a Q-ball is not static, it oscillates slowly, so
that $\varphi_L \to \varphi_L e^{i \omega t}$ and $\varphi_R \to
\varphi_R e^{-i \omega t}$. (The sign is different because the
right-handed quark is described by its left-handed anti-particle and
therefore has opposite baryonic charge.)  This can be accommodated by
changing equations (\ref{massless}) and (\ref{massive}) to
\begin{equation}
\left( \begin{array} {c} \chi_L \\ \chi_R \end{array}\right) = 
\chi_{\rm massless}\left( \begin{array} {c} \sin\alpha e^{-i \omega t} 
\\ -\cos\alpha e^{i \omega t} \end{array}\right)
\label{massless2}
\end{equation}
and
\begin{equation}
\left( \begin{array} {c} \chi_L \\ \chi_R \end{array}\right) = 
\chi_{\rm mass}\left( \begin{array} {c} \cos\alpha e^{-i \omega t}
\\ \sin\alpha e^{i \omega t} \end{array}\right)~.
\label{massive2}
\end{equation}

Now when we project the incoming quark onto the reflected and
transmitted states and then back onto standard left and right-handed
states, the extra phase cancels when the particle retains its
identity and reinforces when the particle changes its
particle--antiparticle identity.  Thus, a quark entering with energy
E will be changed to an anti-quark with energy $E-2\omega$.  The
energy that is lost represents the amount of energy required to raise
the baryon number of the condensate by (2/3), the baryon number of
two quarks.  Such an energy loss is to be expected because $\omega$
can be thought of as a chemical potential of the condensate.

Some final remarks are now in order. In all considerations above we
were taking quarks rather than baryons, which would clearly be more
appropriate at small energies. We believe, however, that the
conclusion about matter-antimatter conversion remains intact in a
realistic case as well, though the computation of  reflection
coefficients is a very difficult task as it requires confinement
effects to be taken into account.  Nevertheless, we expect that a nucleon
can change its baryon number with a large probability on reflection off a
baryonic Q-ball.  Since the lagrangian has no explicit baryon number
non-conservation, this means that the missing baryon number is deposited
into the squark condensate.

\section{Conclusions} 

We have shown that a baryon interacting with a Q-ball can be reflected as
an antibaryon, while the baryon number of a Q-ball is increased by two
units.  This process, which, in general, occurs with a large probability is
the main source of energy release for a Q-ball passing through matter.  In
a separate publication we will reexamine some astrophysical bounds on 
dark matter SUSY Q-balls.

{\bf Acknowledgments.}  The authors thank J.~Morris, E.S.~Phinney, and
P.~Tinyakov for helpful comments.  The work of A.K. and L.L. were supported
in part by the US Department of Energy grant DE-FG03-91ER40662, a NASA ATP
grant NAG5-13399, and by a Faculty Grant from UCLA Council on Research. The
work of M.S. was supported in part by the Swiss Science Foundation.



\end{document}